\documentclass{ptapap}
\usepackage{hyperref}

\author{Klaudia Kowalczyk}[CAMK]
\author{Ewa L. {\L}okas}[CAMK]
\author{Monica Valluri}[UMich]
\affil[CAMK]{Nicolaus Copernicus Astronomical Center,
  Bartycka 18, 00--716 Warszawa, Poland}
\affil[UMich]{Department of Astronomy, University of Michigan,
  1085 S University Ave, Ann Arbor, MI 48109, USA}

\title{Orbit anisotropy of dark matter haloes with Schwarzschild modelling}

\begin{document}

\maketitle

\begin{abstract}

We apply the Schwarzschild orbit superposition method to mock data in order to
investigate the accuracy of recovering the profile of the orbit anisotropy. The mock data come from
four numerical realizations of dark matter haloes with well defined anisotropy profiles.
We show that when assuming a correct mass distribution we are able to determine the anisotropy with
high precision and clearly distinguish between the models.

\end{abstract}

\section{Introduction}
We present the application of the Schwarzschild orbit superposition method
\citep{schwarz_1979} to dark matter haloes, as a first step towards realistic
modelling of orbital structure of dwarf spheroidal galaxies in the Local Group. We assume that the
total mass of the galaxy can be approximated as a spherical dark matter halo and explore the
capabilities of the method in reproducing the underlying orbital anisotropy.

\section{Data}

We used four numerical realizations of stable, spherically symmetric dark matter
haloes of $10^6$ particles each. The models
shared the same density profile: the cuspy NFW \citep{NFW_1997} distribution of virial
mass $M_v=10^9\mathrm{M_\odot}$ and concentration $c=20$ with steeper cut-off at
the virial radius. Our four models differed only in the orbit anisotropy. We considered: an
isotropic model with constant anisotropy parameter $\beta=0$, a radially anisotropic one
with constant $\beta=0.5$ and two models with anisotropy profiles varying with radius, growing
(and decreasing) from $0$ ($0.5$) in the centre to $0.5$ ($0$) at infinity. The models were
generated using the distribution function of \citet{wojtak_2008} and were described in detail
in \citet{gajda_2015} where they are referred to as models C1, C3, I2 and D.

\begin{figure}
\includegraphics[width=\textwidth, trim=-100 34 -100 0, clip]{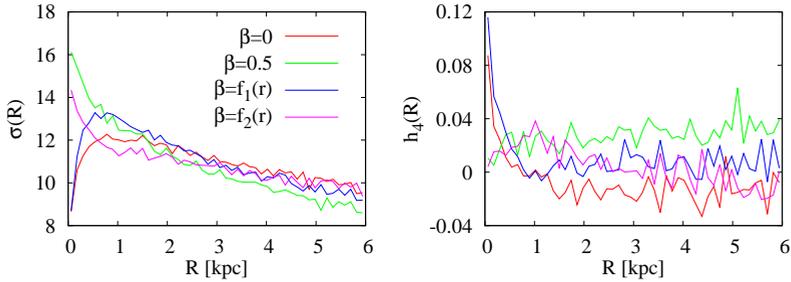}
\caption{The profiles of the velocity dispersion (left panel) and the 4th Gauss-Hermite moment $h_4$
(right panel) for the four haloes with different anisotropy: $\beta=0$ (red), $\beta=0.5$ (green), increasing $\beta$
(blue) and decreasing $\beta$ (magenta).}
\label{fig:sigma_h4}
\end{figure}

We have observed each halo along a random line of sight and binned particles in 50 radial rings spaced
linearly in projected radius up to the distance of $6$ kpc from the centre. In each ring we have stored the velocity
profile $L(v)$ and fitted it with the formula:
\begin{equation}
L(v)=\frac{\gamma\ e^{-(1/2)w^2}}{\sqrt{2\pi}\ \sigma} [1 + h_3H_3(w) + h_4H_4(w)],
\ \ \ \ w=\frac{v-V}{\sigma}
\end{equation}
where $H_3,\ H_4$ are the 3rd and 4th Hermite polynomials, with normalization $\gamma$, the mean
velocity $V$, the velocity dispersion $\sigma$ and the 3rd and 4th Gauss-Hermite moments $h_3,\ h_4$. Such a
fit assumes that $(h_0,\ h_1,\ h_2)=(1,\ 0,\ 0)$. The result of the fitting procedure for $\sigma$
and $h_4$ as a function of radius is presented in Fig.~\ref{fig:sigma_h4}.

\section{Orbit library}

Assuming the density profile matching that of the haloes, we have generated initial conditions for
a large library of 5000 representative orbits. The orbits have been integrated using the public $N$-body
code GADGET-2 \citep{springel_2005} saving in total 2001 points per orbit.
Each orbit has been randomly rotated 200 times around two axes of the
simulation box and combined. In order to extract the observables, we have projected the stuck
orbits along the line of sight and stored them on the same grid as the mock data. The Gauss-Hermite moments have been
calculated using formula (7) from \citet{vdMarel_1993} with the values of $\gamma$, $V$ and
$\sigma$ the same as fitted to the data.

\section{Fitting and results}
We find the best-fitting model by minimizing the $\chi^2$ function over the weights of the orbits
$\gamma_k$, following \citet{rix_1997} and \citet{valluri_2004}. We used the projected mass,
analytically calculated deprojected mass and the Gauss-Hermite moments 0-4. The errors adopted for each
quantity are arbitrarily set to $1\%$ for both masses and $0.01$ for Gauss-Hermite moments.

\begin{figure}[h!!!!!!!!!!!!!!!!!!!!!!!!!]
\includegraphics[width=\textwidth, trim=-100 30 -100 15, clip]{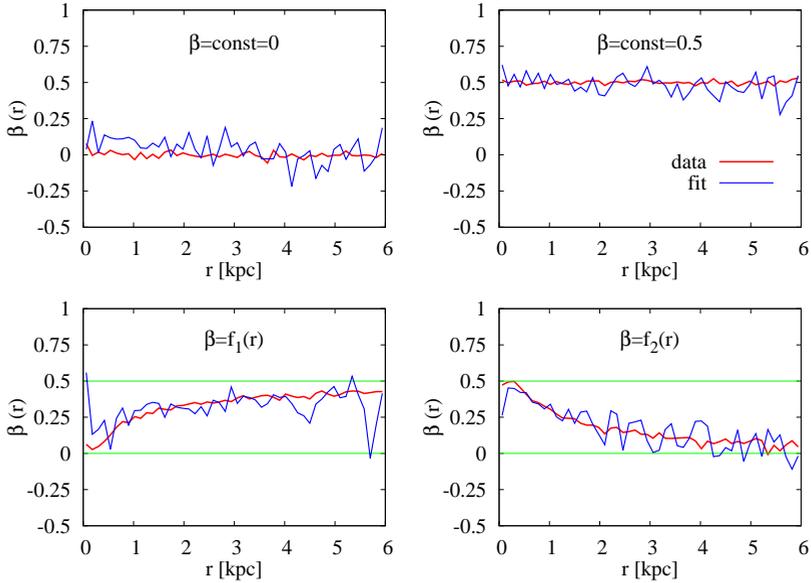}
\caption{The profiles of the anisotropy parameter for the four haloes: from the mock data (red) and
obtained from the fit, i.e. as a result of the Schwarzschild modelling (blue). The
asymptotes of the varying $\beta$ profiles are marked in green.}
\label{fig:beta}
\end{figure}

Similarly to fitted observables, we derive the intrinsic velocity dispersions as sums over the
orbit library weighted with the deprojected masses. We plot the dispersions in terms of the
anisotropy parameter $\beta$ in Fig.~\ref{fig:beta}. We can see that the fitting allows us to recover the
intrinsic anisotropy profiles and distinguish between them.

\section{Summary}
Using numerical realizations of dark matter haloes we have tested the reliability of the
Schwarzschild modelling method in recovering the intrinsic anisotropy of orbits. We have fitted a
library of orbits to the mock data extracted from four haloes of the same mass distribution but
different orbital structure. We have shown that as long as we know the density profile and have
a large sample of tracing particles, by fitting just the projected observables we are able to recover
with high precision the internal orbital properties of the haloes. In our future work we will extend
the application of the method to less idealized conditions.

\acknowledgements{This research was supported in part by the Polish Ministry of Science and
Higher Education under grant 0149/DIA/2013/42 within the Diamond Grant Programme for years
2013-2017 and by the Polish National Science Centre under grant 2013/10/A/ST9/00023.}

\bibliographystyle{ptapap}
\bibliography{ptapap}

\end{document}